%% file: main.tex
\documentclass[twocolumn]{aastex63}

\usepackage{lipsum}
\usepackage{amsmath}
\usepackage{amssymb}
\usepackage{empheq}
\usepackage{mathrsfs}
\usepackage{textcomp}
\usepackage{graphicx}
\usepackage{hyperref}
\usepackage{graphicx}
\usepackage{xspace}
\usepackage[caption=false]{subfig}
\usepackage{multirow}
\usepackage{longtable}
\hypersetup{colorlinks, linkcolor={blue}, citecolor={blue}, urlcolor={blue}}

\newcommand{\kmsmpc}{km\,${\rm s}^{-1}\,{\rm Mpc}^{-1}$\xspace}

\newcommand{\zindex}{{\cal Z}_i}

\begin{document}
\title{Cosmological model insensitivity of local $H_0$ from the Cepheid distance ladder}

\author{S. Dhawan}
\affiliation{The Oskar Klein Centre for Cosmoparticle Physics, Department of Physics, Stockholm University, AlbaNova, 10691 Stockholm, Sweden}

\author{D. Brout}
\affiliation{Department of Physics and Astronomy, University of Pennsylvania, Philadelphia, PA 19104, USA}
\affiliation{NASA Einstein Fellow}

\author{D. Scolnic}
\affiliation{Department of Physics, Duke University, 120 Science Drive,
Durham, NC, 27708, USA}

\author{A. Goobar}
\affiliation{The Oskar Klein Centre for Cosmoparticle Physics, Department of Physics, Stockholm University, AlbaNova, 10691 Stockholm, Sweden}

\author{A. G. Riess}
\affiliation{Space Telescope Science Institute, 3700 San Martin Drive, Baltimore, MD 21218, USA}
\affiliation{Department of Physics and Astronomy, Johns Hopkins University, Baltimore, MD 21218, USA}

\author{V. Miranda}
\affiliation{Steward Observatory, Department of Astronomy, University of Arizona, Tucson, Arizona, 85721, USA}

\begin{abstract}
The observed  tension ($\sim 9\%$ difference) between the local distance ladder measurement of the Hubble constant, $H_0$, and its value inferred from the cosmic microwave background (CMB) could hint at new, exotic, cosmological physics. We test the impact of the assumption about the expansion history of the universe ($0.01<z<2.3$) on the local distance ladder estimate of $H_0$. In the fiducial analysis, the Hubble flow Type Ia supernova (SN~Ia) sample is truncated to $z < 0.15$ and the deceleration parameter ($q_0$) fixed to -0.55. We create realistic simulations of the calibrator and Pantheon samples and account for a full systematics covariance  between these two sets. We fit several physically motivated dark energy models and derive combined constraints from calibrator and Pantheon SNe~Ia and simultaneously infer $H_0$ and dark energy properties. We find that the assumption on the dark energy model does not significantly change the local distance ladder value of $H_0$, with a maximum difference ($\Delta H_0$) between the inferred value for different models of 0.47 \kmsmpc, i.e. a 0.6$\%$ shift in $H_0$, significantly smaller than the observed tension. Additional freedom in the dark energy models does not increase the error in the inferred value of $H_0$. Including systematics covariance between the calibrators, low redshift SNe, and high redshift SNe can induce small shifts in the inferred value for $H_0$.  The SN~Ia systematics in this study contribute $\lesssim 0.8 \%$ to the total uncertainty on $H_0$. 
\end{abstract}

\section{Introduction}
 The Hubble constant describes the present-day expansion rate and sets the absolute distance scale  of the universe. In recent decades, there has been significant progress  in improving the accuracy of measuring $H_0$, with several investigations reporting better than $4\%$ uncertainties in the inferred value of $H_0$ \citep[e.g.][]{2016ApJ...826...56R,2017MNRAS.468.2590S,2019ApJ...882...34F,2019arXiv190704869W,2019arXiv191006306S}.  Estimates of $H_0$ using the local distance ladder approach \citep[e.g.][]{2019ApJ...876...85R,2019arXiv190805625R} are in $\gtrsim 4 \sigma$ tension with the value inferred from the early universe \citep{2018arXiv180706209P}. Furthermore, a completely independent method to measure $H_0$, using time-delay distances to strongly lensed quasars also suggests a high value, exacerbating the tension with the CMB inference to $\gtrsim 5 \sigma$ \citep{2019arXiv190704869W,2019arXiv191006306S}. A summary of the current status of the Hubble tension is provided in \citet{2019arXiv190710625V}.  
 
 The higher value of the local $H_0$ results from any one of five independently determined, geometric distance estimators to calibrate the luminosity of Cepheids in Type Ia supernova (SN~Ia) host galaxies. Independent estimates of $H_0$ from the local, Cepheid distance ladder find no obvious source of systematic error accounting for this discrepancy \citep{Cardona2017,2017MNRAS.471.4946W,2018MNRAS.476.3861F,Follin2017,2017MNRAS.471.2254Z,2018A&A...609A..72D}. Moreover, for quasar time-delay cosmography, \citet{2019arXiv191208027M} find that the inferred value of $H_0$ is robust to sources of systematic uncertainty, e.g. stellar kinematics, line-of-sight effects or assumptions about the lens model. Hence, this observed tension could indicate the presence of exotic physics beyond the standard model \citep[for e.g., see][]{2018JCAP...09..025M,2018JCAP...11..014D,2019arXiv190200534K,2019ApJ...874....4A}. Hence, it is important to examine the impact of various assumptions in the process of inferring the local value of $H_0$ from the different measurement techniques. Here, we analyse the SN~Ia rung of the distance ladder to quantify the impact of the assumption of the cosmological model and sources of systematic uncertainty on the inferred $H_0$ value.

 The magnitude-redshift relation of high-$z$ SNe~Ia was used to discover that the expansion rate of the universe is accelerating, driven by an unknown cosmic component, now termed as dark energy \citep{1998AJ....116.1009R,Perlmutter:1998np}.
 The local value of $H_0$ is estimated using the SN~Ia magnitude-redshift relation \citep{2015ApJ...815..117S}, calibrated with Cepheid variables \citep{2016ApJ...826...56R,2019ApJ...876...85R}. The intercept of the magnitude-redshift relation is computed using SNe~Ia in the nearby ($z < 0.15$) Hubble flow, assuming a fixed value for the deceleration parameter, a dimensionless measure of cosmic acceleration.
 
 In this paper, we analyse the change in the inferred value of local $H_0$ by altering the assumption of the cosmological model describing the expansion history of the universe. There are several viable explanations for the late-time accelerated expansion of the universe \citep[for e.g.;][]{Dhawan2017b,2017ApJ...850..183Z}.  Hence, we simultaneously analyse the SN~Ia magnitude-redshift relation with the Cepheid calibration of the SN~Ia absolute magnitude to test whether $H_0$ is sensitive to the assumption of the model describing the expansion history. 
 We also introduce a new formalism to account for the systematic uncertainties that affect the calibrator and Hubble flow supernovae, motivated for calibrator and $z< 0.15$ Hubble flow SNe in previous studies \citep[e.g.][]{2017MNRAS.471.2254Z,2018MNRAS.476.3861F}.  In \cite{2016ApJ...826...56R}, the SN systematics are treated as variants in the analysis and are not combined in the same way as analyses of the latest high-$z$ SN~Ia samples \citep{2014A&A...568A..22B,2018ApJ...859..101S}.  Here we adopt the formalism used for measuring dark energy properties from high-$z$ samples and extend it to the other rungs of the cosmic distance ladder, so that covariance between the calibrator and Hubble flow SNe distances can be captured for a comprehensive list of systematics and accounted for in the $H_0$ inference.

 We present the methodology and datasets in section~\ref{sec:method}, describe the dark energy models in section~\ref{sec:de_models} and our results in section~\ref{sec:result}. We discuss our findings and conclude in section~\ref{sec:disc}.
 
\section{Methodology and data}
\label{sec:method}
In this section, we describe the datasets and the analysis methodology. For our analysis we use the most recent SN~Ia magnitude-redshift relation from the Pantheon compilation \citep{2018ApJ...859..101S} and the value of the SN~Ia absolute magnitude such corresponding to the fiducial $H_0$ in \citet{2019ApJ...876...85R}. 

For each cosmological model, the distance modulus predicted by the homogeneous and isotropic, flat Friedman-Robertson-Walker (FRW) universe is given by
\begin{equation}
\mu(z; {\boldsymbol \theta} ) = 5\, \mathrm{log_{10}} \left( \frac{D_L}{10\, \mathrm{Mpc}} \right) + 25 \, ,
\label{eq:mu_sne}
\end{equation}
where $z$ is the redshift, ${\boldsymbol \theta}$ are the cosmological parameters (e.g. $\Omega_{\rm M}$, the present day matter density) and $D_L$ is given by

\begin{equation}
D_L = \frac{c(1+z)}{H_0 \sqrt{|\Omega_\mathrm{K}|}}\, \mathrm {sinn}\, \left( \sqrt{|\Omega_\mathrm{K}|} \int^{z}_{0} \frac{dz^{'}}{E(z^{'})} \right) \, ,
\label{eq:lum_dist}
\end{equation}
where  $sinn(X) = {sin(x), x, sinh(x)}$ for closed, flat and open universes and  $E^2(z) = H^2(z)\big/H^2_0$ is the normalised Hubble parameter which describes the expansion history for each model. Throughout this paper, we assume flatness, i.e. $\Omega_{\rm K} = 0$, hence, for each model, the only difference is the expression of $E(z)$. For standard cosmology, $E(z)$ is given by
\begin{equation}
    \frac{H^2(z)}{H_0^2} = \Omega_{\rm M} (1+z)^3 + (1 - \Omega_{\rm M})(1+z)^{3(1+w)}
\label{eq:gen_ez}
\end{equation}

where $\Omega_{\rm M}$ is the present day matter density and $w$ is the equation of state (EoS) of dark energy, which for the standard cosmological model is -1 (hereafter, termed as $\Lambda$CDM).
Observationally, the bias-corrected distance modulus is calculated from the SN~Ia peak apparent magnitude ($m_B$), light curve width ($x_1$) and colour ($c$)
\begin{equation}
\mu_{\rm obs, SN} = m_B - (M_B - \alpha x_1 + \beta c) + \delta_{\rm bias} + \gamma,
\label{eq:obs_distmod}
\end{equation}
where $M_B$ is the absolute magnitude of the SN~Ia,  $\alpha$ and $\beta$ are the nuisance parameters for the width-luminosity and colour-luminosity relations, $\delta_{\rm bias}$ is the 5D distance bias correction following \cite{Kessler_2017} and $\gamma$ is the additional standardization from the correlation between host galaxy stellar mass and SN~Ia intrinsic luminosity following \cite{Conley11}, which was characterized as a step function.
The SN~Ia absolute magnitude is not \emph{a priori} well known, and hence, it requires an independent calibration, e.g. using Cepheid variables. 

In this study, we account for the covariance between the calibrator and Hubble flow SNe~Ia and test how much the inferred $H_0$ changes for different assumptions of the background expansion history. We also compute the contribution of each source of systematic error to the final uncertainty on $H_0$.
We fit the data by minimizing the $\chi^2$ expressed as
\begin{equation}
\chi^2 = \Delta^T C^{-1} \Delta, 
\label{eq:chi}
\end{equation}
where $\Delta = \mu_{\rm th} - \mu_{\rm obs, SN}$ for the SNe~Ia at $z > 0.01$. For the calibrator SNe~Ia, $\Delta = \mu_{\rm obs, SN} - \mu_{\rm Ceph}$. The value of $\mu_{\rm Ceph}$ is chosen such that the SN~Ia absolute magnitude and uncertainty corresponds to the value from \citet{2019ApJ...876...85R}. Therefore, in this study, our fiducial case is the value of M$_B$  that corresponds to $H_0$ from \citet{2019ApJ...876...85R}, and we compute what the shift from this value is under different model assumptions described below in Section~\ref{sec:de_models}. 

Here, $C$ is the complete covariance matrix between the calibrator and the Hubble flow sample of SNe~Ia, described in section~\ref{ssec:covariance}.

We fit the data using a \texttt{python} implementation of the nested sampling software \texttt{MultiNest} \citep{2009MNRAS.398.1601F, 2013arXiv1306.2144F} called \texttt{PyMultiNest} \citep{2014A&A...564A.125B} with 2500 live points and sampling efficiency of 0.8, the recommended value for parameter estimation, and sampling efficiency of 0.3 for computing the evidence.

\subsection{Systematic Uncertainties}
\label{ssec:systematics}
To confidently assess the viability of each cosmological model, we account for numerous sources of systematic uncertainty affecting Hubble flow and calibrator SNe~Ia. The SH0ES team examine systematic shifts in $H_0$ associated with the SN~Ia light curve model, host environments, and the location of a low-z cutoff redshift.  These variants in the analysis are not combined into a full covariance between the calibrators and Hubble flow set, partially because it is difficult to separate statistical fluctuations from systematic shifts in the relatively small sample of SH0ES data.

Here, we improve upon the treatment of systematic uncertainties in the SH0ES analysis by developing the first ever simulations of the sample of calibrators. Such simulations are carried out using the Supernova Analysis (\texttt{SNANA}) software package \citep{Kessler18} which facilitates realistic and fast simulations of SN~Ia datasets. For the calibrator SNe~Ia, we simulate a flat redshift distribution ($0.001<z<0.1$) and we assume survey characteristics and observed fluxes representative of the low-$z$ sample in Pantheon (CfA1-CfA4: \citealt{riess99,jha06,hicken09a,hicken09b,hicken12}; CSP: \citealt{contreras10,Folatelli10,Stritzinger11}). We also simulate each of the high-z rolling surveys identically to Pantheon (SDSS: \citealt{frieman08,kessler09a,sako18}, SNLS: \citealt{Conley11,Sullivan_2011}, PS1: \citealt{rest14,scolnic14a}).

In these simulations we model the impact of 87 different sources of uncertainty and explicitly determine the covariance between the inferred distances to the calibrator and Hubble flow SNe. These systematics are discussed in detail in \cite{2018ApJ...859..101S} and \cite{Brout18-SYS,Brout18-SMP};  here we briefly describe the categories in which they fall.

\textit{Calibration:} We model survey photometric calibration and HST Calspec calibration uncertainties following \cite{2018ApJ...859..101S} and we adopt the SALT2 model calibration systematic uncertainty from \cite{2014A&A...568A..22B}. 

\textit{Host Galaxy Mass:} We model in simulations of SN~Ia host galaxy stellar mass distributions. We adopt the distributions from \cite{Jones_2018} such that 70\% of Hubble flow hosts are high mass (log$_{10}(M_\odot)>10$) and 50\% are high mass for the calibrator sample. We estimate an associated systematic uncertainty by forcing a +0.025 mag shift away from the observed correlation between SN magnitude and host stellar mass.

\textit{z-Bias:} We model the possibility of a small coherent $4\times10^{-5}$ redshift bias as done in \cite{Brout18-SYS} motivated by \cite{Davis_2019}. 

\textit{Intrinsic Scatter Model:} Our nominal analysis assumes the \cite{chotard11} model for intrinsic brightness variations dominated by spectral variations, however we also account for the possibility of the \cite{Guy_2010} model prescribing the majority of intrinsic fluctuations to coherent scatter.

\textit{Milky Way Extinction:} We adopt a global 4\% scaling uncertainty of $E (B - V )_{\rm MW}$ based on the fact that \cite{Schlafly11}, in a re-analysis of
\cite{Schlafly10}, derive smaller values of reddening by 4\%, despite using a very similar SDSS footprint.

\textit{Low-z Sample:} To account for the systematic in modeling of the low-z sample, we vary the outlier cuts from 3.5 to 3 $\sigma$ following \cite{Brout18-SYS}.

\subsection{Computing the covariance}
\label{ssec:covariance}
Accounting for each of the systematics, following \cite{Conley11} and \cite{2018ApJ...859..101S}, we compute a redshift binned systematic covariance matrix. Using BBC \citep{Kessler_2017} fitted distances, for each source of systematic uncertainty (`SYS') we define distances relative to
a nominal analysis (`NOM') as follows:
\begin{equation}
\label{eq:deltamu}
\Delta\langle\mu_{\rm SYS}\rangle_{\zindex} \equiv \langle\mu_{\rm SYS}\rangle_{\zindex}- \langle\mu_{\rm NOM}\rangle_{\zindex} ,
\end{equation}
for redshift bins 
\begin{equation}
\mathcal{Z}(i) = \{ z_{\rm calib}, z_{\rm Pantheon} \} ,
\end{equation}
where $z_{\rm calib} \in \{0, 0.01\}$ is a single bin containing all calibrator SNe~Ia and $z_{\rm Pantheon}$ are the  40 redshift bins from \cite{2018ApJ...859..101S}. For each source of systematic uncertainty, we compute $\langle\mu_{\rm SYS}\rangle_{\zindex}$ by varying that source and re-computing bias corrected distances for both the calibrators and Hubble flow SNe.
\begin{figure*}
    \centering
    \includegraphics[width=.7\textwidth]{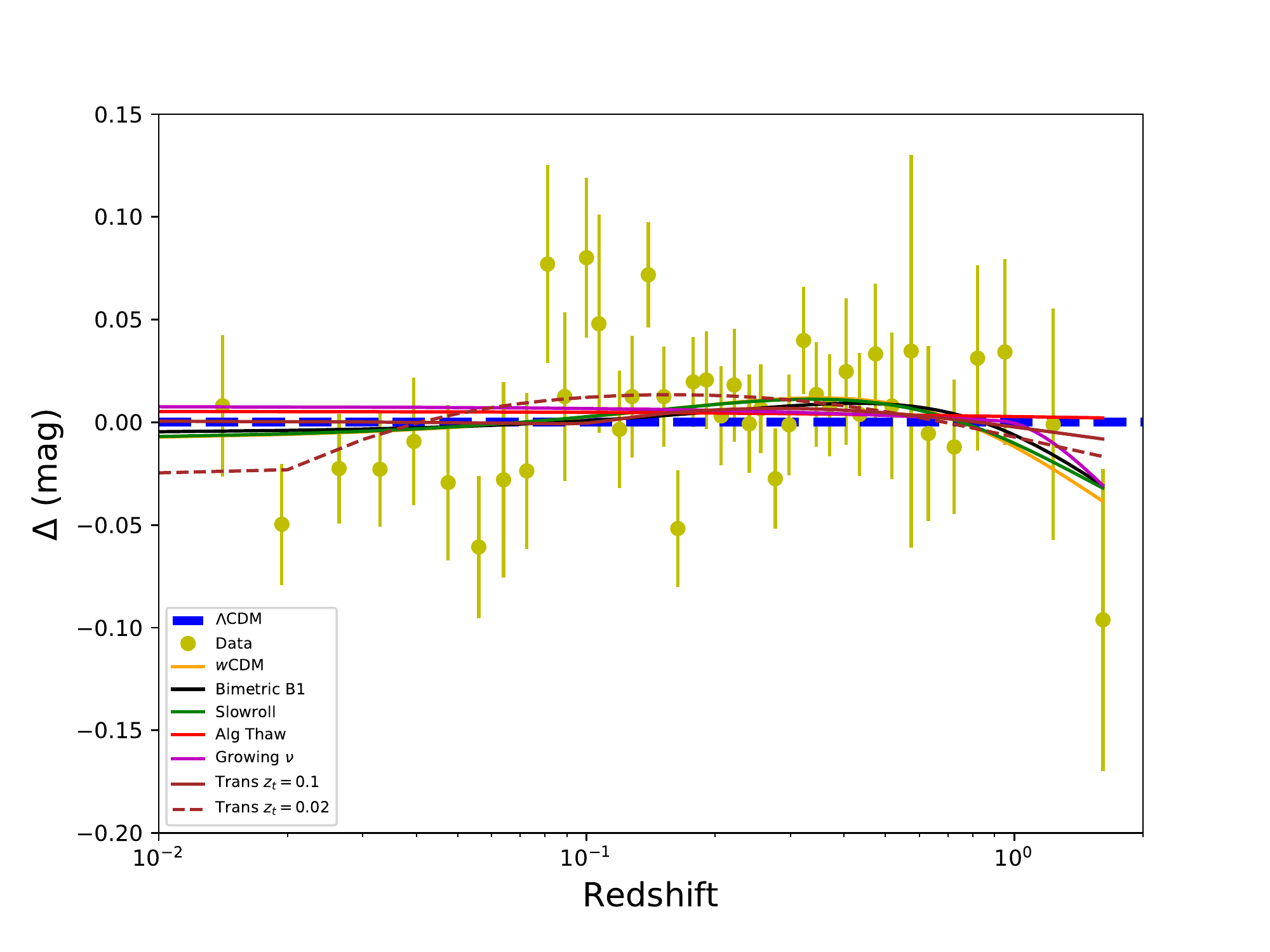}
    \caption{The Hubble residuals as a function of redshift for each dark energy model relative to the best fit $\Lambda$CDM model. The residuals for the data are plotted relative to the best fit $\Lambda$CDM model.}
    \label{fig:model_residuals} 
\end{figure*}

We build our redshift-binned systematic covariance matrix $C_{\rm syst}$ for all sources (${\rm SYS}_k$),
\begin{equation}
\label{eq:csys}
C_{{\zindex}_i{\zindex}_j,{\rm syst}} = \sum_{k=1}^{K=87} \frac{\partial \Delta\langle\mu_{\rm SYS}\rangle_{{\zindex}_i}}{\partial {\rm SYS}_k} ~ \frac{\partial \Delta\langle\mu_{\rm SYS}\rangle_{{\zindex}_j}}{\partial {\rm SYS}_k} ~ \sigma_k^2,
\end{equation}
which denotes the covariance between the $\mathcal{Z}_i^{th}$ and $\mathcal{Z}_j^{th}$ redshift bin summed over the $K$ different sources of systematic uncertainty ($K=87$) with magnitude $\sigma_k$.\\
The covariance matrix used to constrain cosmological models (Eq. 4) is defined as
\begin{equation}
\label{eq:cstatplussyst}
C = C_{\rm stat} + C_{{\rm syst}}
\end{equation}
where $C_{\rm stat}$ is the diagonal matrix of $\sigma_{\mu}^2$ binned in redshift from the publicly available SH0ES and Pantheon samples.

Here we perform both `SH0ES-like' constraints with systematic covariance between the calibrator SNe~Ia and the Hubble flow SNe~Ia in the restricted redshift range ($z<0.15$) as well as full systematic covariance analyses for all SNe~Ia in Pantheon. Our `SH0ES-like' analysis leverages the large SN~Ia statistics of Pantheon however does not include covariance between the calibrator bin ($z_{\rm calib}$) and any bins with $z>0.15$ which greatly reduces systematic uncertainties (hereafter referred to as ``Sys-cutz"). However, because our dark energy models have freedom at all redshifts, we consider our fiducial case with the full systematics covariance matrix without any cuts on redshift.

\input{prior_tab.tex}
\section{Dark Energy Models}
\label{sec:de_models}
Recent studies have shown that several different models of accelerated expansion are a viable explanation of the current data \citep{Dhawan2017b,2017ApJ...850..183Z}. Here, we compile a set of dark energy models with several different physical motivations and define the dimensionless Hubble parameter for each of them. We emphasise that the aim of this study is not to constrain specific models but to analyse a range of different physical explanations for dark energy and their impact on the inferred value of $H_0$.
For each of the models below, the present day matter density $\Omega_{\rm M}$ is a common parameter (except in the model independent case of the cosmographic expansion). For models with additional parameters, we summarize the priors used in our analysis in Table~\ref{tab:prior}. 

The models tested here include a phenomenological extension of $\Lambda$CDM ($w$CDM), a slow-rolling field similar to inflation \citep[one-parameter slow-roll dark energy;][]{2014MNRAS.438.1948S},  a modification to Einstein's general relativity \citep[bimetric gravity;][]{2012JCAP...03..042V,2012JHEP...01..035V,2012JHEP...03..067C,2013JHEP...03..099A}, a dynamical scalar field \citep[Algebraic thawing;][]{2008GReGr..40..329L,2015PhRvD..91f3006L}, a coupling between the neutrino mass and the acceleron field driving accelerated expansion \citep[Growing $\nu$ mass;][]{2007PhLB..655..201W}, and low-redshift dark energy transitions \citep{2009PhRvD..80f7301M}. In addition to these models, which make assumptions about the energy density of the universe, we also test a cosmographic expansion of the Hubble parameter $H(z)$ to the data. We describe these below.

\subsection{One-parameter slow-roll dark energy}
This model is motivated by dynamical behaviour of dark energy.
Recently, \citep{2011MNRAS.416..907G,2014MNRAS.438.1948S} suggest that the simplest dark energy model has the same explanation as inflation, likely a scalar field slowly rolling down its potential. In such a model, dark energy will have a generic equation of state (EoS) and the universe will have a generic dependence of the Hubble parameter on redshift, independent of the potential's starting value and shape.  
The Hubble parameter for this model is given by
\begin{multline}
\frac{H^2}{H_0^2} = \Omega_{\rm M} (1+z)^{3} \\ + (1 - \Omega_{\rm M}) \left[\frac{(1+z)^3}{\Omega_{\rm M}(1+z)^3 + 1 - \Omega_{\rm M}}\right]^{\delta w_0/(1 - \Omega_{\rm M})}
\label{eq:hz_slowroll}
\end{multline}

\vspace{.04in}

\subsection{Bimetric gravity}
This model involves a modification of the equations of general relativity (GR).
Early attempts to modify GR included effectively giving mass to the particle that mediates the gravitational force. It was long believed that massive gravity theories necessarily contained fatal ghost modes \citep{1972PhRvD...6.3368B}. Recently, it was suggested that the inclusion of a second metric and a carefully constructed interaction between the two metrics of the theory could remove the ghost problem \citep{2011PhRvL.106w1101D}. For details on the specific bimetric gravity model tested here, we refer the reader to \citet{2018JCAP...09..025M}. The dimensionless Hubble parameter for this model is
\begin{multline}
\frac{H^2}{H_0^2} = \frac{\Omega_{\rm M} (1+z)^3}{2} +\frac{B_0}{6} \\ + \sqrt{\bigg(\frac{\Omega_{\rm M} (1+z)^3}{2}+ \frac{B_0}{6}\bigg)^2 + \frac{B_1^2}{3}} \, ,
\label{eq:bimetric_eq1}
\end{multline}
with
\begin{equation}
B_0 = 3\Bigg(1 - \Omega_{\rm M} - \frac{B_1^2}{3}\Bigg) \, .
\label{eq:bimetric_eq2}
\end{equation}

\vspace{.04in}

\subsection{Algebraic thawing}
This model belongs to a class of quintessence cosmologies in which the scalar field has a thawing behaviour. Thawing scalar fields that are neither fine-tuned nor have overly steep potentials must initially depart from the cosmological constant behaviour along a specific track in the equation of state phase space, characterised by the form of a slow roll behaviour in the matter-dominated era. The Hubble parameter for this model is given by
\begin{multline}
\frac{H^2}{H_0^2} = \Omega_{\rm M} (1+z)^3 + (1 - \Omega_{\rm M})\,\, \times \\
\mathrm{exp} \Bigg\{ \frac{3(1+w_0)}{\alpha p} \bigg[1 - \bigg(1 - \alpha + \frac{\alpha}{(1+z)^3}\bigg)^{p/3} \bigg]
\Bigg\}\, ,
\label{eq:hz_alg_thaw}
\end{multline}
where $\alpha = 1\big/(1+b)$ and $b = 0.3$ is a fixed constant \citep{2008GReGr..40..329L}.

\subsection{Growing $\nu$ mass} 
Growing neutrino mass models, wherein the mass of the neutrino ($m_\nu$) increases with time and stops the dynamical evolution of the dark energy scalar field are invoked to solve the cosmological coincidence problem, i.e. the problem that the present day matter density and density of $\Lambda$ are similar order of magnitude despite their different dependence on the scale factor \citep{2004JCAP...10..005F,2007PhLB..655..201W}. 

The combined dark sector (scalar field plus neutrinos) energy density (where $a = 1/(1+z)$) is given by 

\begin{equation}
\Omega_{\mathrm{ds}}(a) = \frac{\Omega_{\mathrm{ds}} (a^3) + 2\Omega_\nu (a^{3/2} - a^3)}{1 - \Omega_{\mathrm{ds}} (1 - a^3) + 2\Omega_\nu (a^{3/2} - a^3)}; \,\,  a \textgreater a_t \\
\label{eq:ods_grownu_posttrans}
\end{equation} 
\begin{equation}
\Omega_{\mathrm{ds}}(a) = \Omega_e; \,\,\, a \textless a_t 	
\label{eq:ods_grownu_pretrans}
\end{equation}
where $\Omega_{{\rm ds}} = 1 -\Omega_{\rm M}$\, is the present day dark energy density. The scale factor at which the neutrinos become non-relativistic, is given by preserving continuity between the early and late time terms (i.e. setting equations~\ref{eq:ods_grownu_posttrans} and~\ref{eq:ods_grownu_pretrans} at $a = a_t$). The two free parameters are the early dark energy density $\Omega_{\rm e}$ and the neutrino density $\Omega_\nu$. 
The normalised Hubble parameter for this model is given by 
\begin{equation}
\frac{H^2}{H_0^2} = \frac{\Omega_{\rm M} a^{-3}}{1 - \Omega_{\mathrm{ds}}(a)},
\label{eq:hz_grownu}
\end{equation}

\begin{figure*}
    \centering
    \includegraphics[width=.8\textwidth]{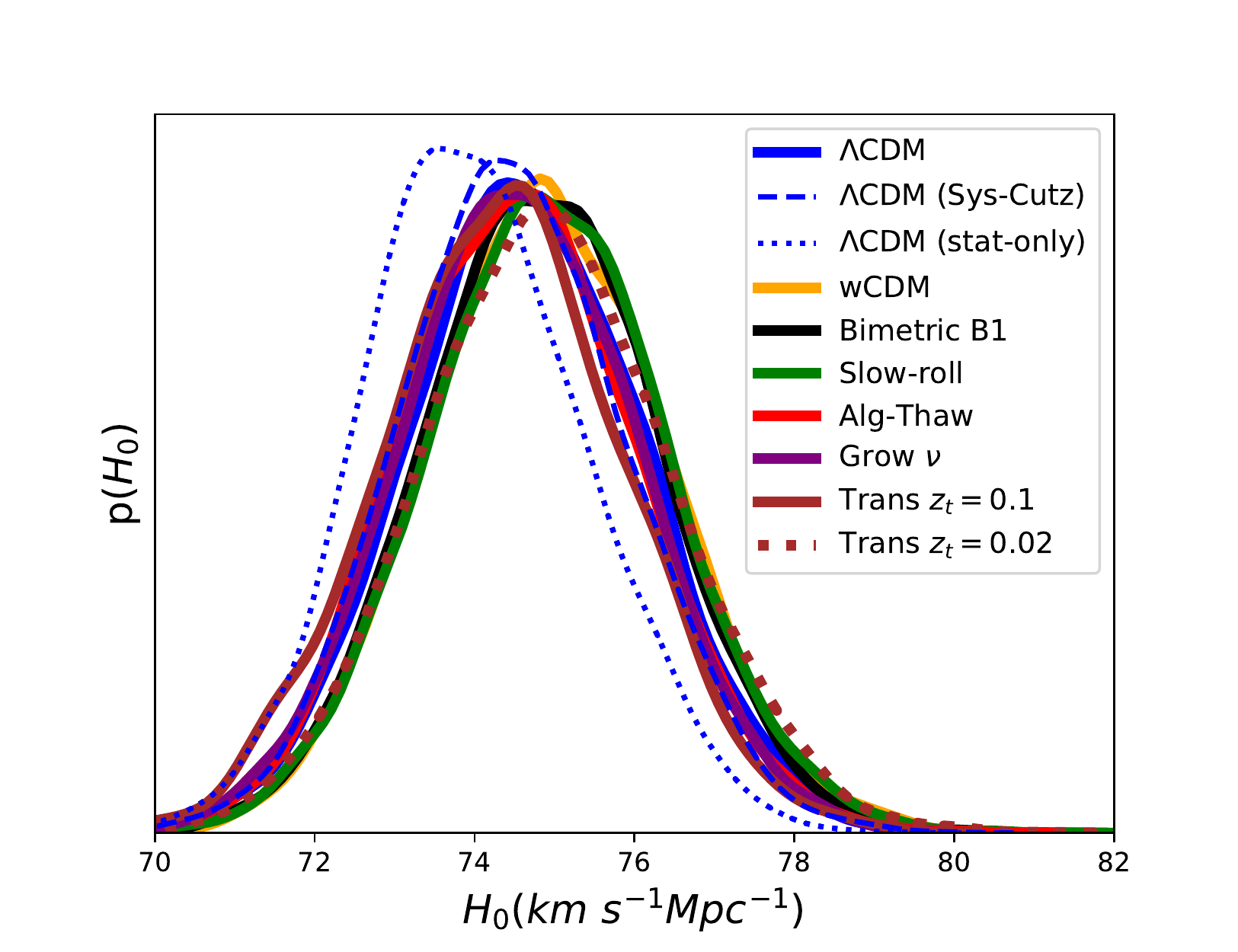}
    \caption{The probability density of $H_0$ for the different cosmological models describing the SN magnitude-redshift relation. The solid lines show the marginalised distribution for $H_0$ for each assumed model and the dotted blue line is the case for the standard $\Lambda$CDM scenario with only statistical uncertainties. The median value and the 1-D marginalised posterior distribution for the different models are very similar (see text for more details). The SN~Ia absolute magnitude is chosen to reproduce the fiducial analysis in \citet{2019ApJ...876...85R}.}
    \label{fig:h0_dist}
\end{figure*}

\subsection{Dark Energy Transitions at Low Redshift}
\citet{2009PhRvD..80f7301M} proposes models with large fluctuations in the dark energy equation of state at low redshifts, typically $z \lesssim 0.02$, that induce step-like transitions in the $H(z)$. Such changes are hidden from constraints coming from Hubble flow SNe~Ia (since even the lowest redshift SNe can typically be at $z > z_t$), but calibrators offer an additional restriction. Step-like responses in $H(z)$ also evade several model-independent constraints, as they often assume smoothness of $H(z)$. 

Phenomenologically, we can write the first Friedmann equation in these models as     
\begin{equation}
    \frac{H^2}{\tilde{H}_0^2} = \tilde{\Omega}_{\rm M}(1+z)^3 + \Bigg[1 + \frac{2 \delta \times \mathcal{S}(z)}{(1-\tilde{\Omega}_{\rm M})\mathcal{S}(0)} \bigg] (1-\tilde{\Omega}_{\rm M})\, ,
\label{eq:trans_de}
\end{equation}
where 
\begin{equation}
    \mathcal{S}(z) \equiv \frac{1}{2}\Bigg[1 - {\rm tanh}\bigg(\frac{z - z_t}{\Delta z}\bigg)\bigg]\, .
\label{eq:transition}
\end{equation}
In this definition, $z_t$, $\Delta z = z_t\big/10$ and $\delta$ are the position, width and amplitude of the transition respectively. In particular, $\delta = 0$ corresponds to the standard $\Lambda$CDM. Hence, this model has three free parameters at a given fixed $z_t$: $\tilde{H}_0$, $\tilde{\Omega}_{\rm M}$ and $\delta$. The observed Hubble constant is $H_0 = \tilde{H}_0 \sqrt{1 + 2\delta} \neq \tilde{H}_0$. 

In this work, we either assume $z_t = 0.02$ or $z_t = 0.1$; the former case motivated by \citet{2009PhRvD..80f7301M} themselves and the latter case motivated so we can use all the low-$z$ SNe at $z \approx 0.1$. The adopted priors for $\tilde{H}_0$ and $\tilde{\Omega}_{\rm M}$ are equivalent to the priors shown on Table~\ref{tab:prior} for $H_0$ and $\Omega_{\rm M}$. The difference in notation highlights the fact that $\tilde{H}_0$ and $\tilde{\Omega}_{\rm M}$ cannot be interpreted as the observed Hubble constant and the dark matter density at redshift zero in these models (see~\citealt{2009PhRvD..80f7301M}). 

\subsection{Cosmographic expansion}
\label{ssec:cosmograph}
Along with the different dark energy models described above we also look into a more model-independent method, by expanding the expression for $H(z)$ as a Taylor series,  linearly in $z$ in a cosmographic approach \citep[for e.g.;][]{2019PhRvL.122f1105F,2019MNRAS.486.2184M,2019arXiv190512000A,2019arXiv190907986A,2020PhRvR...2a3028C}. This approach has been used previously in inverse distance ladder estimates of $H_0$ \citep[e.g.][]{Bernal2016,2019MNRAS.483.4803L,2019PhRvL.122f1105F}.

Expanding $H(z)$ we get,

\begin{equation}
    H(z) = H_0 \big(1+\mathcal{B}_1z +  \mathcal{B}_2z^2 + \mathcal{B}_3z^3\big)
\label{eq:cosmograph}
\end{equation}
where $\mathcal{B}_1 = 1 + q_0$, 
2$\mathcal{B}_2 = j_0 - q_0^2$ and $6\mathcal{B}_3 = 3 q_0^3 + 3q_0^2 - j_0(3+4q_0) - s_0$,  $j_0$ is the cosmological jerk and $s_0$ is the snap parameter.

\section{Results}
\label{sec:result}
In this section we present the results of fitting the different dark energy models described in section~\ref{ssec:DE_mod} to the combined calibrator and Hubble flow SNe~Ia. We also discuss the impact of the systematics covariance matrix on the inferred value of $H_0$ in section~\ref{ssec:sys_err}.

\subsection{Dark energy model fits}
\label{ssec:DE_mod}

\begin{figure}
    \centering
    \includegraphics[width=.45\textwidth]{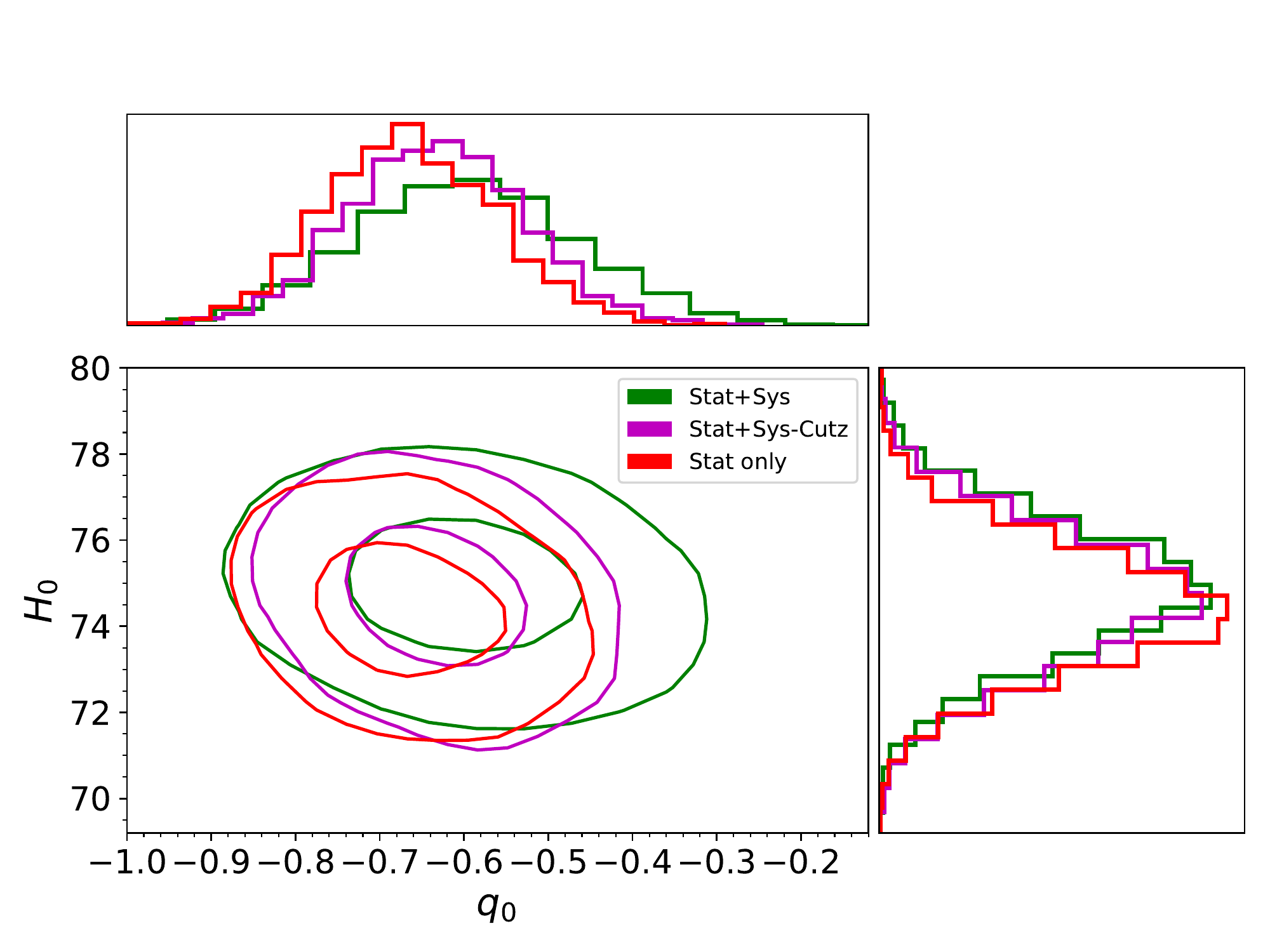}
    \caption{The joint posterior distribution on $H_0$ and $q_0$ for the cosmographic expansion of the dimensionless Hubble parameter as a function of redshift (Equation~\ref{eq:cosmograph}) for the case with the complete systematics covariance matrix (green), only $z < 0.15$ SNe~Ia having systematic uncertainties (magenta) and the case with only statistical uncertainties (red).}
    \label{fig:cosmograph}
\end{figure}
We fit all the models listed in Section~\ref{sec:de_models} to the combined calibrator and Hubble flow SN~Ia data described in section~\ref{sec:method}. For each of the non-standard dark energy models, we plot the distances corresponding to the best fit cosmological parameters relative to the best fit standard $\Lambda$CDM case in Figure~\ref{fig:model_residuals}. The models have very similar residuals to $\Lambda$CDM, except for small differences at higher redshift. We find that these models have best fit values close to their $\Lambda$CDM limit. For each model, we also report the logarithm of the Bayesian evidence, $Z$, such that the $\Delta ln\, Z$ can be used for model comparison. The Bayesian evidence is the average likelihood over the prior region \citep[see][for details]{2017arXiv170101467T}, expressed as, 
\begin{equation}
Z = \int L \pi d \theta
\label{eq:evidence}
\end{equation}
where $L$ is the likelihood, $\pi$ is the prior and  $\theta$ is the set of parameters. The prior values for each model parameter are presented in Table~\ref{tab:prior}. 
The resulting $H_0$ distribution assuming each of the dark energy models is shown in Figure~\ref{fig:h0_dist}. We find that the $H_0$ value inferred  is very insensitive to the assumption of the cosmological model. At its most extreme, the difference between the $H_0$ value for the highest and lowest case is 0.47 \kmsmpc (see Figure~\ref{fig:h0_dist}). This corresponds to a maximum shift of 0.6$\%$ in the $H_0$ value, significantly smaller than the uncertainty on $H_0$ or the discrepancy of the distance ladder value with the value inferred from the early universe. We note that the dark energy transition model with $z_t = 0.02$  has the highest improved $\chi^2$ relative to the standard $\Lambda$CDM scenario,  however, it has a slightly lower Bayesian evidence, owing to the model having more free parameters. This is possibly due to the slight offset of the lowest redshift bins relative to the higher-z bins, which is due to the difference in the intrinsic scatter model assumed \citep[see][for details]{2018ApJ...859..101S}.

\input{h0vals_tab.tex}

We fit the cosmographic expansion described in section~\ref{ssec:cosmograph}. This allows us to simultaneously fit for $H_0$ and the parameters defining accelerated expansion (i.e. $q_0$, $j_0$, $s_0$), independent of assumptions on the underlying cosmological model. 
The resulting posterior distribution of $H_0$, $q_0$ is presented in Figure~\ref{fig:cosmograph}. We find that for the cosmographic approach, $H_0$ is consistent with the values derived assuming different dark energy models in section~\ref{sec:de_models}. The inferred value of $q_0$ is $-0.59 \pm 0.14$, which is consistent with the expected value for the standard cosmological model (i.e. $q_0 = -0.55$). 
We emphasise here that the low value of ln\,$Z$ for the cosmographic expansion is due to the large uniform prior on the model parameters. We use a large prior region to explore a large parameter space for deriving the posterior distribution, which makes the ln\,$Z$ small despite the model being a good fit to the data.
\input{sys_err_tab.tex}
\subsection{SN systematic error contribution}
\label{ssec:sys_err}
As described in section~\ref{ssec:covariance}, we account for the covariance between the calibrator and Hubble flow samples. The SN~Ia systematic uncertainties contribute $\sim 0.8\%$ to the total uncertainty on $H_0$. 
We find that the case with covariance between only the $z < 0.15$ Hubble flow SNe and the calibrator sample returns very similar results to the case with the full covariance (see blue dashed line compared to the solid blue line in  Figure~\ref{fig:h0_dist}). This is also true for the cosmographic expansion, shown in Figure~\ref{fig:cosmograph} (magenta and green histograms).

A summary of the individual contributions to the final systematic error budget is shown in Table~\ref{tab:sys_err}. 
The largest sources of systematic uncertainty are from photometric calibration and the assumed model of intrinsic scatter. This is similar to the {$w$-error} budgets of high redshift supernova cosmology analyses (JLA: \citealt{2014A&A...568A..22B}, Pantheon: \citealt{2018ApJ...859..101S}, DES: \citealt{Brout18-SYS}). However specifically for $H_0$ analyses, the contribution of the host galaxy mass systematic is amplified when the distributions of masses are not consistent between the calibrator hosts and the Hubble flow hosts as we have examined here. We also find that both an offset of $E(B-V)_{\rm MW}$ values and a potential local void bias in Hubble flow redshifts contribute insignificantly to the final uncertainty on $H_0$. The total systematic error contribution from Hubble flow SNe~Ia to the final $H_0$ uncertainty is 0.58 \kmsmpc, which is roughly the same size as the statistical uncertainty, and a final error on $H_0$ of $\sim$ 1.4 \kmsmpc,  in agreement with \cite{2019ApJ...876...85R}.

In addition to examining the contribution of each systematic to the final $H_0$ uncertainty, we also examine shifts in the recovered central value for $H_0$. We find a shift of 0.73 in $H_0$ from the inclusion of systematics covariance between the calibrators and Hubble flow SNe. The shift in the inferred value of $H_0$ from each individual source of systematic error is summarised in Table~\ref{tab:sys_err}. We find similar results for either intrinsic scatter model.  To understand the origin of such shifts in the central value of $H_0$, we generated a mock covariance matrix with zero covariance between the calibrator and Hubble flow SNe. With this mock covariance, we find $\Delta H_0 \simeq 0$ relative to the $\Lambda$CDM model with only statistical uncertainties. We simulated another mock covariance in which the calibrator SNe have the same covariance value with each of the Hubble flow SN bins and find that again $\Delta H_0 \simeq 0$.  We find that shifts in $H_0$ from SN covariance arise from non-zero and non-constant covariance between the calibrator SN bin and the individual Hubble flow SN bins with the direction of the shift depending on the signs of the covariance.  The shift we find is a realistic value for the Pantheon SN sample and SH0ES calibrators.


\section{Discussion and Conclusions}
\label{sec:disc}

Here, for the first time, we present a combined analysis of the high-redshift ($0.01 < z < 2.3$) SN~Ia and the Cepheid distances to nearby SN~Ia host galaxies to compute $H_0$ using a combined SN covariance matrix. The covariance matrix includes several sources of uncertainty to account for correlation between the SNe~Ia in the calibrator and Hubble flow samples  \citep[for e.g., see][]{2018ApJ...859..101S,Brout18-SYS}. We find that the SN~Ia systematics contribute $\lesssim 0.8 \%$ to the total uncertainty budget for $H_0$. 

Interestingly, we find that inclusion of covariance between the Cepheid calibrator SNe and the various Hubble flow SNe can induce small shifts in $H_0$ ($\lesssim 0.75$\,\kmsmpc) relative to a statistical only analysis. Our study finds that these shifts arise specifically from differing covariance between the Cepheid calibrator and the Hubble flow bins. Although these shifts are smaller than the present errors in $H_0$ they will be important to address in future distance ladders which seek to approach a precision of $\sim$ 1\%.

Several studies in the literature have found the local value of $H_0$ to be robust to different sources of systematic uncertainty, e.g. the statistical inference model, sample variance, Cepheid systematics and using near infrared data for SNe~Ia \citep{Cardona2017,2017MNRAS.471.4946W,2018MNRAS.476.3861F,Follin2017,2017MNRAS.471.2254Z,2018A&A...609A..72D}. From our comprehensive study of systematics, we find the largest contribution to the $H_0$ uncertainty is from the photometric calibration and the assumed model for SN~Ia intrinsic scatter, whereas we find little contribution from potential redshift measurement biases as well as little contribution from MW extinction likely due to the fact that extinction offsets are absorbed in the SN color terms.

In the fiducial analysis of the local distance ladder, the deceleration parameter, $q_0$ is fixed to -0.55, corresponding to the standard cosmological model. We tested what the impact of the assumption of the dark energy model is on the inferred value of $H_0$.
Using a diverse range of physically motivated models for dark energy, we find that the maximum difference in the inferred $H_0$ is 0.47 \kmsmpc, a shift of 0.6$\%$. The best fit constraints on the expansion history for each of these models to be close to  $\Lambda$CDM. While the dark energy models tested here do not shift $H_0$ significantly from the fiducial value, models with oscillating parameters \citep{2019ApJ...875...34B} could be a possible candidate to shift the value of $H_0$, however, they have already been ruled out by current data. 
Furthermore, we analysed dark energy models with low- and ultra low-redshift transitions in the equation of state \citep{2009PhRvD..80f7301M}. We find that in both cases of a low-$z$ transition at $z_t=0.1$ and ultra low-$z$ transition at $z_t =0.02$, there is no significant shift in the central value of $H_0$. 

 We report the log of the ratio of Bayesian evidences for each non-standard dark energy model relative to the model with the highest evidence, i.e. $\Lambda$CDM. While most models are indistinguishable relative to $\Lambda$CDM, there is moderate evidence ($\Delta {\rm ln}\, Z > 3$) disfavouring the algebraic thawing model. We note that the prior ranges assumed for the model parameters are narrow. Even in a more extreme case of a narrower range, using U[-1, 0] as the prior on $w_0$, we get only a slightly improvement in the evidence relative to the $\Lambda$CDM case. Previously, samples of SNe~Ia, e.g. JLA \citep{2014A&A...568A..22B} could not distinguish between models like algebraic thawing and standard $\Lambda$CDM with SNe~Ia alone, i.e. without combining with complementary cosmological probes \citep[e.g.][]{Dhawan2017b}. This demonstrates the importance of reducing SN~Ia systematic uncertainties for improving dark energy model selection.

We also computed $H_0$ for a model independent approach using a cosmographic expansion of the Hubble parameter as a function of redshift. For this approach, we find no significant shift in $H_0$ and $q_0 = -0.59 \pm 0.14$ which is consistent with the value of $q_0$ in standard cosmology. Comparing the case with the complete systematics covariance matrix to the case with only covariance between the calibrator and Hubble flow SNe with $z < 0.15$, we find no significant difference in the inferred $H_0$. The SN~Ia systematics contribute $\lesssim 0.8\%$ to the uncertainty on $H_0$.
We, therefore, conclude that the assumption about the model describing accelerated expansion does not significantly change the inferred value of $H_0$.  


\acknowledgments
We would like to thank Nikki Arendse for helpful discussions and comments. SD and AG acknowledge support from the Swedish Research Council. DB acknowledges support for this work was provided by NASA through the NASA Hubble Fellowship grant \#HST-HF2-51430.001 awarded by the Space Telescope Science Institute, which is operated by Association of Universities for Research in Astronomy, Inc., for NASA, under contract NAS5-26555. VM was supported by NASA ROSES ATP 16-ATP16-0084 and NASA ADAP 16-ADAP16-0116 grant. DB was partially supported by DOE grant DE-FOA-0001781.

\bibliographystyle{aasjournal}

\bibliography{h0_cosmo}
\end{document}

%% file: prior_tab.tex
\begin{table*}
\centering
\caption{Priors on the free parameters for the models tested in this study.}
\begin{tabular}{|l|c|c|}
\hline\hline
Model Parameter & Prior & Model\\
\hline
$\Omega_{\rm M}$ & U[0, 1] & All\\
$w$ & U[-2, 2] & $w$CDM \\
$w_0$ & U[-1, 1] & Algebraic thawing\\
$p$ & U[-4, 4] & Algebraic thawing \\
$B_1$ & U[0, 6] & Bimetric gravity \\	
$\delta w_0$ & U[-2, 1] & One-parameter Slow-roll dark energy \\
$\Omega_{\rm e}$ & U[0, 0.25] & Growing $\nu$ mass\\
$\Omega_{\nu}$ & U[0, 0.4] & Growing $\nu$ mass\\
$\delta$ & U[-0.4, 0.6] & Dark Energy Transitions at Low Redshift \\
$H_0$ & U[50, 100] & All\\
$q_0$ & U[-5, 5] & Cosmographic expansion \\
$j_0$ & U[-5, 5] & Cosmographic expansion \\
$s_0$ & U[-10, 10] & Cosmographic expansion \\
\hline
\end{tabular}
\label{tab:prior}
\end{table*}

%% file: h0vals_tab.tex
\begin{table*}
\centering
\caption{Inferred $H_0$, ln\,$Z$ and $\Delta$ ln\,$Z$  for different dark energy models governing the expansion history of the universe. A higher value of $\Delta$ ln\,$Z$ indicates that the model is more disfavoured relative to standard cosmology.}
\begin{tabular}{|l|c|c|c|r|r|}
\hline\hline
Model & $H_0$ & ln\,$Z$ & $\Delta$ ln\,$Z$ & {\bf $\chi^2$} & {\bf \# Param.}\\
& (\kmsmpc) & & &\,\, &  \\
\hline

$\Lambda$CDM & $74.62 \pm 1.48$ & -32.22 & $\ldots$ & {\bf 37.89} & {\bf 3}\\
$\Lambda$CDM (stat-only) & $73.94 \pm 1.38$ & -38.68 & $\ldots$ & {\bf 47.46} & {\bf 3}\\
$w$CDM & $74.93 \pm 1.51$ & -33.88 & 1.66 & {\bf 36.25} & {\bf 4}\\
Bimetric-Linear and Quadratic (BQ) & $74.88 \pm 1.49$ & -32.23 & 0.01 & {\bf 36.20} & {\bf 4}\\
Slowroll & $74.84 \pm 1.50$ & -33.77 & 1.55 & {\bf 36.56} & {\bf 4}\\
Alg-Thaw & $74.53 \pm 1.50$ & -35.73 & 3.51 & {\bf 37.98} & {\bf 5}\\
Grow-$\nu$ & $74.56 \pm 1.51$ & -34.26 & 2.04 & {\bf 37.24} & {\bf 5}\\
Trans: $z_t = 0.1$ & $74.37 \pm 1.50$ & -35.40 & 3.18 & {\bf 37.40} & {\bf 4} \\
Trans: $z_t = 0.02$ & $74.90 \pm 1.57$ & -33.88 & 1.66 & {\bf 35.20} & {\bf 4}\\ 
&&& &\,\, &\\
Cosmographic Expansion & $74.88 \pm 1.54$ & -37.23 & 5.10 & {\bf 36.30} & {\bf 5}\\
\hline
\end{tabular}
\label{tab:h0_val}
\end{table*}

%% file: sys_err_tab.tex
\begin{table}
\centering
\caption{The contribution to the final $H_0$ uncertainty, for the case assuming $\Lambda$CDM cosmology from each source of systematic error in the covariance matrix.}
\resizebox{.42\textwidth}{!}{\begin{tabular}{|l|c|r|}
\hline\hline
Source & $\sigma(H_0)$ & $\Delta H_0$\footnote{Shift relative to the case with only statistical uncertainties. The positive value of the shift indicates a higher value of $H_0$, and negative lower.} \\
& (\kmsmpc) & (\kmsmpc)  \\
\hline

 Photometric Calibration & 0.39 & 0.21\\
 Intrinsic Scatter Model & 0.27 & 0.20 \\
 Host Mass Distribution &  0.26 & 0.12\\
 SALT2 & 0.16 & 0.16\\
 Low-z Modeling/Outliers & 0.13 & -0.05 \\
 MW Extinction & 0.03 & 0.05\\
 z-Bias ($5\times10^{-5}$) & 0.02 & 0.04\\
\hline
Total & 0.58 & 0.73\\
\hline
\end{tabular}
}
\label{tab:sys_err}
\end{table}